\documentclass[conference,final,twoside]{IEEEtran}

\usepackage{flushend}
\usepackage{amssymb}
\usepackage{url}
\usepackage{ifthen}
\usepackage{graphicx}
\usepackage{fancyvrb}
\usepackage{times}
\usepackage{hyperref}

\newenvironment{STDdescription}
               {\list{}{\labelwidth 0pt \itemindent-\leftmargin
                        }}
               {\endlist}
\renewcommand*\descriptionlabel[1]{\hspace\labelsep
                                   \normalfont\bfseries #1}

\makeatletter
\newenvironment{btHighlight}[1][]
{\begingroup\tikzset{bt@Highlight@par/.style={#1}}\begin{lrbox}{\@tempboxa}}
{\end{lrbox}\bt@HL@box[bt@Highlight@par]{\@tempboxa}\endgroup}

\newcommand\btHL[1][]{%
  \begin{btHighlight}[#1]\bgroup\aftergroup\bt@HL@endenv%
}
\def\bt@HL@endenv{%
  \end{btHighlight}%
  \egroup
}
\newcommand{\bt@HL@box}[2][]{%
  \tikz[#1]{%
    \pgfpathrectangle{\pgfpoint{1pt}{0pt}}{\pgfpoint{\wd #2}{\ht #2}}%
    \pgfusepath{use as bounding box}%
    \node[anchor=base west, outer sep=0pt,inner xsep=1pt, inner ysep=0pt, rounded corners=2pt, minimum height=\ht\strutbox+1pt,#1]{\raisebox{1pt}{\strut}\strut\usebox{#2}};
  }%
}
\makeatother

\newboolean{LONGVERSION}
\setboolean{LONGVERSION}{false}

\ifthenelse{\boolean{LONGVERSION}}{
\usepackage{varioref}
}{
}

\VerbatimFootnotes
\IEEEoverridecommandlockouts
\begin{document}

\title{BARR-C:2018 and MISRA C:2012: \\
       Synergy Between the \\
       Two Most Widely Used C~Coding Standards}
\author{\IEEEauthorblockN{Roberto Bagnara\textsuperscript{*}\thanks{%
{}\textsuperscript{*} While Roberto Bagnara
is a member of the \emph{MISRA C Working Group} and of
ISO/IEC JTC1/SC22/WG14, a.k.a.\ the \emph{C Standardization Working Group},
the views expressed in this paper are his and his coauthors' and should
not be taken to represent the views of either working group.}}
\IEEEauthorblockA{BUGSENG and University of Parma\\
Parma, Italy\\
Email: roberto.bagnara@bugseng.com}
\and
\IEEEauthorblockN{Michael Barr}
\IEEEauthorblockA{Barr Group\\
Germantown, Maryland, U.S.A.\\
Email: mbarr@barrgroup.com}
\and
\IEEEauthorblockN{Patricia M.\ Hill}
\IEEEauthorblockA{BUGSENG\\
Parma, Italy\\
Email: patricia.hill@bugseng.com}}

\maketitle

\begin{abstract}
The \emph{Barr Group's Embedded~C Coding Standard} (BARR-C:2018, which
originates from the 2009 Netrino's Embedded~C Coding Standard) is,
for coding standards used by the embedded system industry, second only
in popularity to MISRA~C.  However, the choice between MISRA~C:2012
and BARR-C:2018 needs not be a hard decision since they are
complementary in two quite different ways.  On the one hand, BARR-C:2018
has removed all the incompatibilities with
respect to MISRA~C:2012 that were present in the previous edition
(BARR-C:2013). As a result, disregarding programming style,
BARR-C:2018 defines a subset of C that, while preventing a significant
number of programming errors, is larger than the one defined by MISRA
C:2012. On the other hand, concerning programming style, whereas MISRA
C leaves this to individual organizations, BARR-C:2018 defines a
programming style aimed primarily at minimizing programming errors.
As a result, BARR-C:2018 can be seen as a first, dramatically useful
step to C language subsetting that is suitable for all kinds of
projects; critical projects can then evolve toward MISRA~C:2012
compliance smoothly while maintaining the BARR-C programming style.
In this paper, we introduce BARR-C:2018, we describe its relationship
with MISRA~C:2012, and we discuss the parallel and serial adoption of
the two coding standards.
\end{abstract}

\section{Introduction}
\label{sec:introduction}

The C~programming language is still, after half a century from its
inception, among the most used programming languages overall\footnote{%
  Source: TIOBE Index for December 2019,
  see \url{https://www.tiobe.com/tiobe-index/}.}
and the most used one for the development of embedded systems
\cite{BarrGroupSurvey2018,VDCSurvey2011}.
The reasons for such success are deeply rooted in compelling industry
requirements and have been discussed elsewhere (see, e.g.,
\cite{BagnaraBH18,BagnaraBH19}).
Among such requirements are language size, stability and an evolution
path that ensures backward compatibility.

Faithfulness of C to its original spirit is also a cause of problems.
As discussed in \cite{BagnaraBH18,BagnaraBH19}, each strong point of
C comes with a corresponding weakness:
\begin{itemize}
 \item
   the ease of writing efficient compilers for almost any
   architecture, the existence of many compilers by different vendors
   and the fact that C is defined by an ISO standard are the reasons
   why the language is not fully defined;
\item
  the objective of easily obtaining efficient code with no hidden costs
  has been achieved, in C, also by ruling out all run-time error checking;
\item
  easy access to the hardware comes with the risk of inadvertently
  corrupt the program state;
\item
  language terseness opens the door to misunderstanding and abuse
  of the language that easily results into program that are obscure
  and unsuitable for code reviews.
\end{itemize}

The potential impact of the mentioned weak points of C is of course
higher for critical applications.
One of the pragmatic solutions adopted by industry
to mitigate this problem  is called \emph{language subsetting}:
critical applications are not programmed in unrestricted C,
but in a subset where the potential of committing possibly
dangerous mistakes is reduced.
This is mandated or highly recommended by all functional safety
standards, such as
IEC 61508 \cite{IEC-61508:2010} (industrial, generic),
ISO~26262 \cite{ISO-26262:2011} (automotive),
CENELEC~EN~50128 \cite{CENELEC-EN-50128:2011} (railways),
RTCA DO-178B/C \cite{RTCA-DO-178C} (aerospace) and
FDA's \emph{General Principles of Software Validation} \cite{FDA02}
(medical devices).

Of course, coding in a safer subset of C is not enough to guarantee
correctness.  However:
\begin{enumerate}
\item
  The restriction to a language subset where not fully defined behavior
  and problematic features are banned or severely regulated ``can considerably
  help the efficiency and precision of the static analysis''
  \cite{CousotCFMMMR07}.
  In the case of C, the restrictions posed on features like
  unions, pointer casts and backward gotos
  can be exploited in the design of static analysis tools
  \cite{BagnaraBH18}.
\item
  Properly designed language subsets have a strong emphasis on
  code readability: code reviews combined with static analysis
  and the automatic enforcement of sound coding guidelines by
  means of high-quality tools are the basis of most effective
  defect removal strategies \cite{JonesB11}.
\end{enumerate}
Concerning the second point, code readability is also influenced
by coding practices that go beyond language subsetting (see, e.g.,
\cite{dosSantosG18}).
These have to do with code layout, naming of program entities,
contents of source and header files, and use of comments.

A recent survey conducted among embedded system software professionals
\cite{BarrGroupSurvey2018} found that,
setting aside proprietary coding standards,
\emph{MISRA~C} \cite{MISRA-C-2012-Revision-1}
is the most widely used coding standard,
and \emph{BARR-C} \cite{BARR-C-2018} is the next most widely used.%
\footnote{The survey described in \cite{BarrGroupSurvey2018} does not
  come with statistical significance guarantees, especially as far as
  the randomness of the surveyed group is concerned.  However, we
  believe the sample size of 1,703 firmware designers is more than
  respectable and the survey findings are matched by the daily
  professional experience of the present authors.}
The survey found that, together, these coding standards
were the primary basis of the project-specific coding standards followed
by more than 40\% of respondents.

BARR-C was not designed to compete against MISRA~C:
they are in fact compatible and complementary.
In this paper, after a brief introduction of
MISRA~C:2012 \cite{MISRA-C-2012-Revision-1},
we introduce
BARR-C:2018 \cite{BARR-C-2018}
by highlighting its relationship with
MISRA~C:2012, and we discuss the parallel and serial adoption of
the two coding standards.

The plan of the paper is the following:
Section~\ref{sec:c-traps-and-pitfalls} recalls some well-known
C~language traps and pitfalls;
Section~\ref{sec:misra-c} introduces the MISRA project and MISRA~C;
Section~\ref{sec:barr-c-2018-introduction} introduces BARR-C;
Section~\ref{sec:barr-c-2018-language-subsetting-guidelines} presents
all the language-subsetting guidelines of BARR-C:2018
and their relationship with the MISRA~C:2012 guidelines;
Section~\ref{sec:barr-c-2018-stylistic-guidelines} presents
all the stylistic guidelines of BARR-C:2018;
Section~\ref{sec:adoption-of-barr-c-2018-and-misra-c-2012} presents
scenarios for the successful adoption of both coding standards;
Section~\ref{sec:conclusion} concludes.

\section{C Traps and Pitfalls}
\label{sec:c-traps-and-pitfalls}

There are two main categories of issues that pose problems to the use of C
in the development of critical systems:
\begin{enumerate}
\item
  the language is not fully defined;
\item
  the language can easily be abused to write obscure code
  that is resistant to reviewing activities.
\end{enumerate}
We briefly review these categories in the following sections.

\subsection{C Is Not Fully Defined}

In this paper we refer to the 1999 version of the ISO C standard
\cite{ISO-C-1999-consolidated-TC3}, which is the latest version
of the standard supported by MISRA~C:2012 and BARR-C:2018.
Nonetheless, the contents of this section applies to all versions
of the ISO C standard.

For the reasons mentioned previously, the C standards leave
several aspects of the language not fully defined.
There are four classes of not fully defined behaviors
\begin{STDdescription}
\item[implementation-defined behavior:]
\emph{unspecified behavior where each implementation documents how the choice is made}
\cite[Par.~3.4.1]{ISO-C-1999-consolidated-TC3};
e.g., the sizes and precise representations of the standard integer types;
\item[locale-specific behavior:]
\emph{behavior that depends on local conventions of nationality, culture, and
language that each implementation documents}
\cite[Par.~3.4.2]{ISO-C-1999-consolidated-TC3};
e.g., character sets and how characters, numbers, dates and times
are displayed;
\item[undefined behavior:]
\emph{behavior, upon use of a non-portable or erroneous program construct or
of erroneous data, for which this International Standard imposes
no requirements}
\cite[Par.~3.4.3]{ISO-C-1999-consolidated-TC3}; e.g., attempting to
write a string literal constant or shifting an expression
by a negative number or by an amount greater than or equal to the width
of the promoted expression;
\item[unspecified behavior:]
\emph{use of an unspecified value, or other behavior where this
International Standard provides two or more possibilities and imposes
no further requirements on which is chosen in any instance}
\cite[Par.~3.4.4]{ISO-C-1999-consolidated-TC3}; e.g., the order in which
actual parameters of function calls are evaluated.
\end{STDdescription}
In the sequel, we will collectively refer to these not fully defined behaviors
as ``non-definite behaviors.''

Setting aside locale-specific behavior, whose aim is to avoid some
nontechnical obstacles to adoption,
it is important to understand the intimate connection between non-definite
behavior and the relative ease with which optimizing compilers
can be written.  In particular, C data types and operations can be
directly mapped to data types and operations of the target
machine.  This is the reason why the sizes and precise representations
of the standard integer types are implementation-defined: the implementation
will define them in the most efficient way depending on properties
of the target CPU registers, ALUs and memory hierarchy.

Overflow on signed integer types is undefined behavior because the C~standard
allows three different representations of signed integers: two's complement,
ones' complement and sign-magnitude.  For one's complement and sign-magnitude,
it is implementation-defined whether the \emph{negative zero} bit pattern
is a \emph{trap representation}.\footnote{Trap representations are particular
  object representations that do not represent values of the object type.
  Simply reading a trap representation is undefined behavior.
  For instance, in a memory architecture with
  explicit parity bits or other error detection/correction codes,
  a representation with a faulty error detection/correction code
  can be a trap representation.}
The C compiler can thus assume signed integer overflow cannot happen
and omit all checks for overflows.

Attempting to write on string literal constants is undefined behavior
because they may reside in read-only memory and/or may be merged and shared:
for example, a program containing \verb|"String"| and \verb|"OtherString"|
may only store the latter and use a suffix of that representation to represent
the former.  So, if the hardware traps attempts to write read-only memory,
an unspecified hardware exception may take place, otherwise the program
might end up changing, due to sharing, more than one string literal.
Again, the C compiler can thus assume the program will never try to
write on a string literal constant.

The reason why shifting an expression by a negative number
or by an amount greater than or equal to the width of the promoted expression
is undefined behavior is due to two factors:
\begin{enumerate}
\item
  Allowing shifting by a negative number of bit positions, an operation
  that is usually not supported in hardware, would require a test, a jump
  and a negation.
\item
  Allowing shifting by an amount greater than or equal to the width
  of the promoted expression not only would be pointless: it would
  require extra machine instructions on those architectures where
  the shift count is reduced by masking in order to reduce the maximum
  execution time of the shift instructions.  For instance, on IA-32
  (Intel Architecture, 32-bit) only the 5 low-order bits are retained
  by a preliminary masking operation.
  As a result, shifting a 32-bit register to the left by 32 positions
  does not result in 0, as one would expect, but in the register being
  unchanged (the 5 low-order bits of 32 are 0, and shifting by 0
  positions is a no-op).
\end{enumerate}
So, for ease of implementation of the compiler
and speed of the generated code, C leaves this behavior undefined.

If a program relies on undefined or unspecified behaviors, then
its semantics is not defined, that is, it is not possible to assign
any meaning to it.
All programs do rely on implementation-defined behavior, so their
semantics can be defined only with reference to the used implementation
of the language \cite{Bagnara19TR}.
Reliance on implementation-defined behavior is an obstacle both for
portability and for understandability of the programs (there is
such a variety of implementation-defined behaviors ---112 in C99---
that most of them are unknown to the majority of programmers).

\subsection{C Can Be Difficult To Read}

There are many features of C that, if not properly used, can
impact program readability and understandability:
\begin{itemize}
\item the preprocessing phase;
\item a generous offer of operators with nontrivial and easily forgotten
  precedence and associativity rules;
\item a generous offer of control-flow mechanisms, some of which
  are characterized by a very complex semantics
  (\verb+goto+, \verb+switch+, \verb+for+,
  \verb+break+, \verb+continue+,
  \verb+setjmp+/\verb+longjmp+, \dots);
\item implicit conversions governed by quite intricate rules;
\item two kinds of comment markers with a nontrivial interaction between
  themselves and with \emph{line splicing} (i.e., splitting logical lines
  into multiple physical lines using trailing backslashes as line-continuation
  markers).
\end{itemize}

Remarkable examples of unreadable code are provided by the winners of
\emph{The International Obfuscated C~Code Contest}, a contest
running without interruption since 1984, which awards a prize
to the most obscure/obfuscated C~programs that respects a few basic rules.%
\footnote{\url{http://www.ioccc.org/}, last accessed on March 15, 2020.}

As already mentioned, ensuring code readability and understandability
is crucial for the effectiveness of code reviews and has an obvious
impact on other program properties such as maintainability.

\section{MISRA C}
\label{sec:misra-c}

This paper is concerned with MISRA C:2012~\cite{MISRA-C-2012}.
This is the latest in a series of
standards for the C language that have resulted from the MISRA
project.
Starting in 1990 with the mission of providing
world-leading best practice guidelines for the safe and secure
application of both embedded control systems and standalone software
\cite{BagnaraBH19}, the project published, in November 1994,
``Development guidelines for vehicle based software''~\cite{MISRA-1994}
prescribing the use of
``a restricted subset of a standardized structured language.''

For this reason, the MISRA consortium began work on
the MISRA C guidelines: at that time Ford and Land Rover
were independently developing proprietary guidelines for vehicle-based
C software and it was recognized that a common activity would
be more beneficial to industry.
The first version of the MISRA~C guidelines was published
in 1998~\cite{MISRA-C-1998} and received significant industrial
attention.

In 2004,
MISRA published an improved version of the C guidelines \cite{MISRA-C-2004}
for which the intended audience was extended to include
\emph{all} industries that develop C software for use in high-integrity/critical
systems.
Due to the success of MISRA~C and the fact that C++
is also used in critical contexts, in 2008, MISRA published  a similar set
of \emph{MISRA~C++} guidelines~\cite{MISRA-CPP-2008}.

Both MISRA~C:1998 and MISRA~C:2004 target the 1990 version of the
C standard~\cite{ISO-C-1990}.
The latest version, MISRA~C:2012, published in 2013~\cite{MISRA-C-2012},
supports both
C99 \cite{ISO-C-1999} as well as C90
(in its amended and corrected form sometimes referred to as C95
\cite{ISO-C-1995}).\footnote{The MISRA~C Working Group
  is currently working, among other things \cite{Banks16},
  at adding support for C11 \cite{ISO-C-2011}
  and C18 \cite{ISO-C-2018}.}
With respect to previous versions, MISRA~C:2012 covers more language
issues and provides a more precise specification of the guidelines
with improved rationales and examples.
MISRA~C:2012 is the most authoritative language subset for the
C programming language.

\begin{figure*}[p]
\begin{center}
\includegraphics[width=13.74cm]{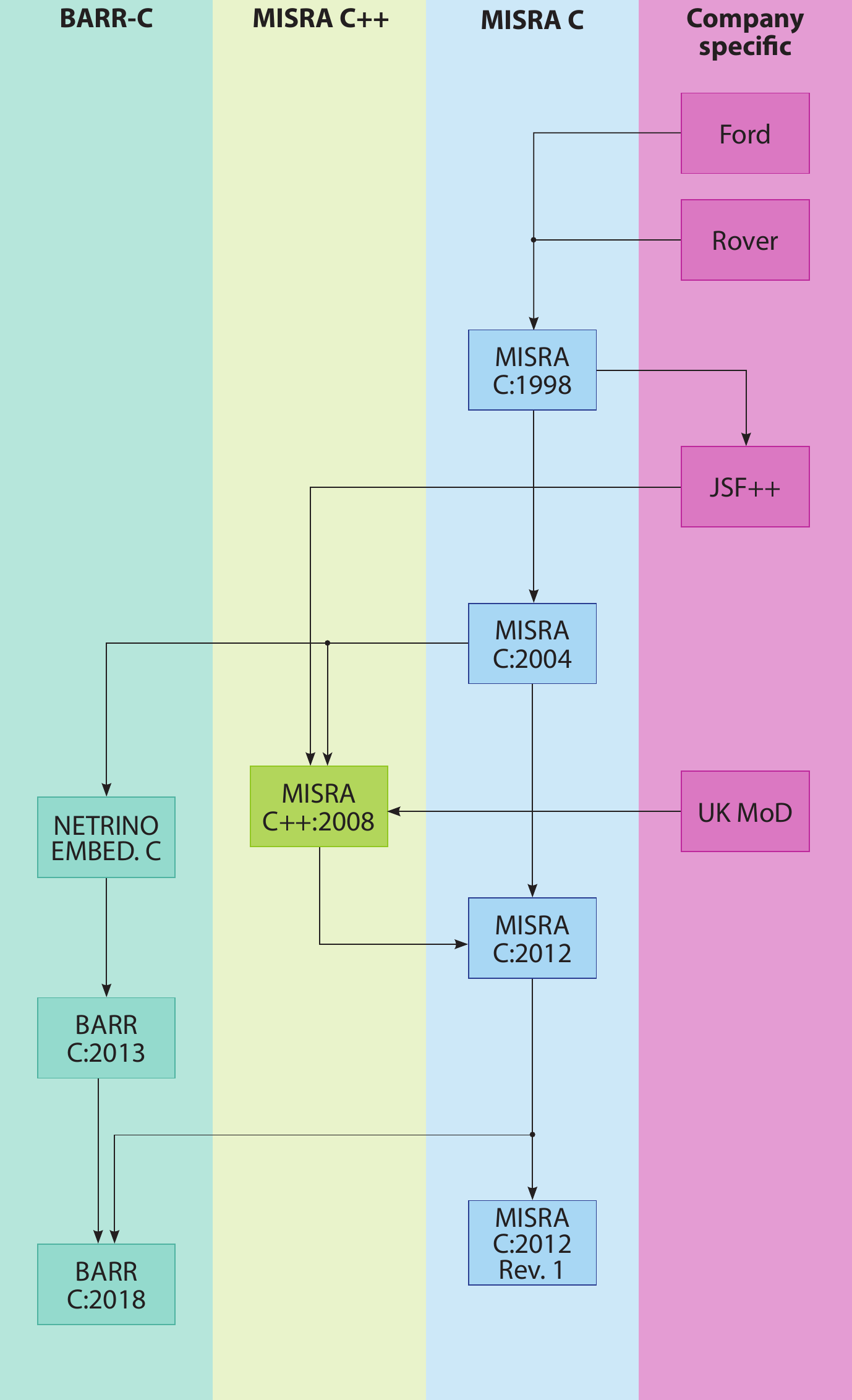}
\caption{Origin and history of MISRA C and BARR-C}
\label{fig:misra-c-history}
\end{center}
\end{figure*}

MISRA~C, in its various versions, influenced all publicly-available
coding standards for C and C++ that were developed after MISRA~C:1998.
Figure~\ref{fig:misra-c-history} shows part of the relationship
and influence between the MISRA C/C++ guidelines and other sets of
guidelines.  It can be seen that MISRA~C:1998 influenced
Lockheed's ``JSF Air Vehicle C++ Coding Standards for
the System Development and Demonstration Program'' \cite{JSF-CPP-2005},
which influenced
MISRA~C++:2008, which, in turn, influenced MISRA~C:2012.
The activity that led to MISRA~C++:2008 was also encouraged
by the UK Ministry of Defence which, as part of its
\emph{Scientific Research Program}, funded a
work package that resulted in the development of a ``vulnerabilities document''
(the equivalent of Annex~J listing the various behaviors in ISO~C,
which is missing in ISO~C++, making it hard work to identify them
and to ensure they are covered by the guidelines).
Moreover, MISRA~C deeply influenced NASA's
``JPL Institutional Coding Standard for the C Programming Language''
\cite{NASA-JPL-C-2009} and several other coding standards (see, e.g.,
\cite{CERT-C-2016,IPA-ESCR-C-2014}), including the BARR-C
coding standards that will be described in the next sections
\cite{BARR-C-2018}.

The MISRA C guidelines are concerned with aspects of C that impact on
the safety and security of the systems, whether embedded or standalone:
they define ``a subset of the C language in which the opportunity to
make mistakes is either removed or reduced'' \cite{MISRA-C-2012}.
The guidelines ban critical non-definite behavior and constrain
the use of implementation-defined behavior and compiler extensions.
They also limit the use of language features that can easily be
misused or misunderstood.  Overall, the guidelines are designed to
improve reliability, readability, portability and maintainability.
We assume the reader has some general familiarity
with MISRA~C:2012 \cite{MISRA-C-2012-Revision-1}: we recommend
reading \cite{BagnaraBH19} if that is not the case.

There are two kinds of MISRA C guidelines:
\begin{STDdescription}
\item[Directive:]
a guideline where the information concerning compliance
is not fully contained in the source code and
requirements, specifications, design, etc.,
may have to be taken into account.
Static analysis tools may be able to assist in checking compliance.
\item[Rule:]
a guideline where information concerning compliance
is fully contained in the source code.
Discounting undecidability, static analysis tools should, in principle,
be capable of checking compliance.
\end{STDdescription}

MISRA~C has been designed to be used
within the framework of a documented development process where
justifiable non-compliances will be authorized and recorded
as \emph{deviations}.
To facilitate this, each MISRA C guideline has been assigned a category.
\begin{STDdescription}
\item[Mandatory:]
C code that complies to MISRA C must comply with every
mandatory guideline; deviation is not permitted.
\item[Required:]
C code that complies to MISRA C shall comply with every
required guideline; a formal deviation is required
where this is not the case.
\item[Advisory:]
these are recommendations that should be followed as far as is
reasonably practical;
formal deviation is not required,
but non-compliances should be documented.%
\footnote{MISRA~Compliance:2016~\cite{MISRA-Compliance-2016}, which is
  optional for MISRA~C:2012 but will become mandatory starting
  from the next version of MISRA~C,
  allows these guidelines to be downgraded to ``Disapplied''.}
\end{STDdescription}
Every organization or project may choose to treat any required guideline
as if it were mandatory and any advisory guideline as if it were required
or mandatory.

Each MISRA~C rule is marked as \emph{decidable} or \emph{undecidable}
according to whether answering the question ``Does this code comply?''
can be done algorithmically.
Rules are marked `decidable'
whenever compliance depends only on compile-time (static) properties
such as the types of the objects or the names and the scopes of identifiers.
Conversely, rules are marked `undecidable' whenever violations depend on
run-time (dynamic) properties such as the value contained in a modifiable
object or whether control reaches a particular point.%
\footnote{Most interesting program properties
such as whether a program can lead to a division by zero,
a buffer overflow or a memory leak are undecidable.}
The majority of the MISRA~C guidelines are decidable,\footnote{Out of a
  total of 173 guidelines, only 36 rules and 4 directives involve undecidable
  program properties \cite{BagnaraBH18}.}
and thus compliance can be checked by algorithms that:
\begin{itemize}
\item
  do not need nontrivial approximations of the value of program objects;
\item
  do not need nontrivial control-flow information.
\end{itemize}
Of course, these algorithms can still be very complex.
For instance, the nature of the translation process of the C~language,
which includes a preprocessing phase, is a source of complications:
the preprocessing phase must be tracked precisely, and compliance
may depend on the source code before preprocessing, on the source code after
preprocessing, or on the relationship between the source code before and
after preprocessing.

MISRA C rules are also classified as \emph{single translation unit} or
\emph{system} according to the amount of code that needs to be
analyzed in order to check compliance.  If a rule is marked `single
translation unit' then compliance can be determined by checking each
translation unit independently.  On the other hand, if a rule is
marked `system', then, to decide the compliance of code in a
specific unit, all the source code in the program (or, in some cases,
project) may need to be checked.

\section{BARR-C:2018: Introduction}
\label{sec:barr-c-2018-introduction}

The history of the \emph{Barr Group's Embedded C Coding Standard}
---BARR-C for short---
started with the publication, in 2009, of the
\emph{Netrino's Embedded~C Coding Standard} \cite{Netrino-C-2009}.
This coding standard, as well as its subsequent versions,
was specifically designed to reduce the number
of programming defects in embedded software as well as
improving maintainability and portability.
\emph{Netrino's Embedded~C Coding Standard} was renamed
\emph{Embedded~C Coding Standard} and released, in 2013,
as a freely downloadable PDF document \cite{BARR-C-2013}.
The next and current version, BARR-C:2018 \cite{BARR-C-2018} has been
improved by ensuring that BARR-C's guidelines can be combined
with MISRA-C:2012's guidelines without conflicts.
Figure~\ref{fig:misra-c-history} shows the various versions of BARR-C
in the larger context where they were developed.

As far as the compatibility between BARR-C:2018 and MISRA~C:2012 is concerned,
the objectives declared in \cite{BARR-C-2018} are:
\begin{enumerate}
\item
  BARR-C-2018 guidelines that define a subset of the C~programming language
  should never be more restrictive than the MISRA~C:2012 guidelines.
  In other words, the subset of the C~language defined by MISRA-C:2012
  should itself be a subset of the subset defined by the BARR-C:2018
  guidelines.
\item
  BARR-C-2018 guidelines that place stylistic limitations on programmers
  (such as restricting code formatting or the names of some identifiers)
  do not contradict the MISRA C guidelines.  In other words,
  BARR-C:2018 includes a C~style guide that is complementary to MISRA C,
  which does not make any recommendations related purely to style.
\end{enumerate}
As we will see in
Section~\ref{sec:barr-c-2018-guidelines-without-correspondence},
objective number~1 has not been fully achieved.

In compiling the BARR-C coding standard, guidelines were selected
for their ability to minimize defects \cite{BARR-C-2018}:
\begin{quote}
\itshape
When it was the case that one rule had the ability to prevent more
defects from being made by programmers than an alternative rule for a
similar aspect of coding, that more impactful rule was chosen.  For
example, the stylistic rules for when and where to place curly braces
were selected on the basis of their ability to reduce bugs across a
whole program.
\end{quote}

BARR-C:2018 does not make a clear-cut distinction between
guidelines such that information about compliance is in the code
and the language implementation ---\emph{rules} in MISRA~C parlance---
and guidelines that require further information ---\emph{directives}
in MISRA~C:2012
\cite{MISRA-C-2012-Revision-1}.
The \emph{Enforcement} section of each guideline description
in BARR-C:2018 provides some indication about the use of (existing) tools
and code reviews for compliance verification.
In this paper, we adopt the MISRA~C view in presenting
BARR-C:2018.  In particular:
\begin{itemize}
\item
  we refer to generic BARR-C:2018 rules as ``guidelines'';
\item
  we refer to BARR-C:2018 rules whose compliance only depends
  on the source code and the used language implementation as
  ``rules'';
\item
  we refer to the remaining BARR-C:2018 rules as ``directives''.
\end{itemize}

BARR-C:2018 guidelines will be introduced using the following format:
\begin{STDdescription}
\item[C.S.I (G{[$\star$][!]})] \emph{Headline}
  (relationship with MISRA~C:2012, if any)
\end{STDdescription}
where
\begin{itemize}
\item
  \textbf{C}, \textbf{S} and \textbf{I} are the chapter, section
  and item letter that uniquely identify the guideline;
\item
  \textbf{G} is either \textbf{D}, for directives, or \textbf{R}, for rules;
\item
  the optional $\star$ symbol flags guidelines that,
  according to~\cite{BARR-C-2018},
  are objectively expected to reduce the number of defects
  (38 out of 143 guidelines of BARR-C:2018 are marked as such);
\item
  the optional \textbf{!} symbols flags guidelines that are dubbed
  ``bug-killing'' in \cite{BarrS14};\footnote{While \cite{BarrS14} mentions
    10 rules, one of them, ``Rule \#5'' in \cite{BarrS14} is the combination
    of guidelines \textbf{2.1.b} and \textbf{2.1.c}
    of \cite{BARR-C-2018}. There are thus
    11 \textbf{!}-tagged guidelines in the present paper.}
\item
  \emph{headline} is a brief summary of the guideline.
\end{itemize}
Please note that in several cases the \emph{headline} has been conceived
just for this paper and for illustrative purposes only: the reader should
check the real, full guideline text in \cite{BARR-C-2018}.

All but 5 BARR-C:2018 rules can be checked at the level of the single
translation unit: the exceptions are \textbf{1.8.a}, \textbf{1.8.b},
\textbf{4.2.c}, \textbf{7.2.a}, and \textbf{8.4.b}.
All but 3 of them are decidable: the exceptions are
\textbf{1.8.b}, \textbf{7.2.a}, and \textbf{8.4.b}.

\section{BARR-C:2018 Language Subsetting Guidelines}
\label{sec:barr-c-2018-language-subsetting-guidelines}

In this section we present the language subsetting guidelines in
BARR-C:2018.  We have 64 such guidelines (out of 143),
which we divide into four categories:
\begin{enumerate}
\item[A.] guidelines with an exact match in MISRA~C:2012;
\item[B.] guidelines with a non-exact match in MISRA~C:2012;
\item[C.] guidelines with related guidelines in MISRA~C:2012;
\item[D.] guidelines without correspondence in MISRA~C:2012.
\end{enumerate}
Each category is introduced in the four following sections.

\subsection{BARR-C:2018 Guidelines with an Exact Match in MISRA~C:2012}

10 guidelines of BARR-C:2018 have an (almost) exact match in MISRA~C:2012.
They are:

\begin{STDdescription}
\item[1.1.d (R)] Do not use preprocessor directive \verb+#define+ to
  define macros with the same name as a keyword (MISRA~C:2012 Rule~20.4).
\item[1.4.a (R$\star$)] Do not rely on C's operator precedence rules
  (MISRA~C:2012 Rule~12.1).
\item[1.8.a (R$\star$)] All declarations and definitions of variables
  or functions at file scope must have internal linkage unless
  external linkage is required
  (MISRA~C:2012 Rule~8.7).
\item[2.1.c (D$\star$!)]  Code shall never be commented out
  (MISRA~C:2012 Dir~4.4).
\item[6.2.c (R)] There must be only one \verb+return+ statement exiting a
  function and this must be at the end of the function
  (MISRA~C:2012 Rule~15.5).\footnote{This guideline is controversial,
    but the single point of exit is explicitly
    required by IEC~61508 \cite{IEC-61508:2010} and
    ISO~26262 \cite{ISO-26262:2011}.}
\item[6.2.e (R$\star$!)] Any object or function declaration or
  definition with internal linkage must include the \verb+static+
  storage class specifier (MISRA~C:2012 Rule~8.8).
\item[6.2.f (R)] Each parameter shall be explicitly declared and
  meaningfully named (MISRA~C:2012 Rule~8.2).
\item[6.3.a (D$\star$!)] Use function-like macros only when they are not
  replaceable by (inline) function calls (MISRA~C:2012 Dir~4.9).
\item[7.2.a (R$\star$)] All variables shall be initialized before use
  (MISRA~C:2012 Rule~9.1).
\item[8.2.d (R)] Any \verb+if+ statement with an \verb+else if+
  clause shall end with an \verb+else+ clause (MISRA~C:2012 Rule~15.7).
\end{STDdescription}

\subsection{BARR-C:2018 Guidelines with a Non-Exact Match in MISRA~C:2012}

27 guidelines of BARR-C:2018 have a non-exact match in MISRA~C:2012.
They are:

\begin{STDdescription}
\item[1.1.a (R)] All programs shall be written to comply with the C99 version
  of the ISO C Programming Language Standard
  (MISRA~C:2012 Rule~1.1).
  The MISRA guideline is less restrictive in that it allows also C90.
\item[1.1.b (R)] Whenever a C++ compiler is used, appropriate compiler
  options shall be set to restrict the language to the selected version
  of ISO C
  (MISRA~C:2012 Rule~1.1).
\item[5.2.a (D$\star$!)] Use the \verb+typedef+s provided by \verb+<stdint.h>+
  instead of the basic integer types (MISRA~C:2012 Dir~4.6).
  The MISRA guideline is more general in that: it covers also the
  floating-point types; it covers also C90, for which it allows
  the use of type names different from those provided by \verb+<stdint.h>+.
\item[1.3.a (R$\star$!)] Braces shall always surround the bodies of \verb+if+,
  \verb+else+, \verb+switch+, \verb+while+, \verb+do+, and \verb+for+
  (MISRA~C:2012 Rule~15.6, for iteration and selection statements,
  and Rule~16.1, for switch statements).
\item[1.7.c (R)] Possibly avoid all uses of the \verb+goto+ keyword.
  If goto is used it shall only jump to a label declared later in the same
  or an enclosing block (MISRA~C:2012 Rule~15.1, advisory ban of \verb+goto+,
  and Rule~15.2, required limitation to forward jumps).
\item[1.8.b (R$\star$!)] The \verb+const+ keyword shall be used
  whenever appropriate (MISRA~C:2012 Rule~8.13, advisory restriction
  to \verb+const+-qualified pointees).
\item[2.1.b (R$\star$!)] Comments shall never contain the preprocessor tokens
  \verb+/*+, \verb+//+, or \verb+\+
  (MISRA~C:2012 Rule~3.1 covers \verb+/*+ and \verb+//+,
  Rule~3.2 covers \verb+\+).
\item[4.2.b (R)] Each header file shall contain a preprocessor guard against
  multiple inclusion.
  (MISRA~C:2012 Dir~4.10, except that BARR-C:2018 Rule \textbf{4.2.b}
  also requires
  a comment after the closing \verb+#endif+).
\item[5.3.a (R)] Bit-fields shall not be defined within signed integer types
  (MISRA~C:2012 Rule~6.1 allows bitfields that are explicitly signed/unsigned).
\item[5.3.b (R$\star$!)] None of the bitwise operators (i.e., \verb+&+,
  \verb+|+, \verb+~+, \verb+^+, \verb+<<+, and \verb+>>+) shall be used
  to manipulate signed integer data
  (MISRA~C:2012 Rule~10.1 covers more than bitwise operators and conditions
  are more general).
\item[5.3.c (R$\star$!)] Signed integers shall not be combined with unsigned
  integers in comparisons or expressions.  In support of this, decimal
  constants meant to be unsigned should be declared with a `\verb+u+ suffix
  (MISRA~C:2012 Rule~10.4 covers the first part of this rule,
  MISRA~C:2012 Rule~7.2 covers the second part).
\item[6.1.a (R$\star$)] No procedure shall have a name that is a keyword of any
  standard version of the C or C++ programming language
  (MISRA~C:2012 Rule~21.2 covers only the keywords of the applicable
  C~standard).
\item[6.1.b (R$\star$)] No procedure shall have a name that overlaps
  a function in the C~Standard Library
  (MISRA~C:2012 Rule~21.2 covers all kinds of identifiers and macro names,
  not just function names).
\item[6.1.c (R)] No procedure shall have a name that begins with an underscore
  (MISRA~C:2012 Rule~21.2 prevents declaration of any identifier with
  a reserved name, and this includes those beginning with underscore).
\item[6.1.d (R)] No procedure name shall be longer than 31 characters
  (MISRA~C:2012 Rule~5.1 requires external identifiers to be distinct
  which, for a generic C99 compiler, implies different external procedure names
  shall differ in the first 31 characters).
\item[6.2.d (R$\star$)] A prototype shall be declared for each public function
  in the module header file
  (MISRA~C:2012 Rule~8.5 covers this rule, but it applies also to objects
  and requires one and only one declaration per function/object).
\item[6.3.b (R$\star$)] If parameterized macros are used, the following
  restrictions apply:
  (i) macro body surrounded in parentheses;
  (ii) macro parameters surrounded in parentheses;
  (iii) macro parameters used no more than once;
  (iv) do not include control transfer statements
  (MISRA~C:2012 Rule~20.7 covers the same issues addressed by points~i and~ii;
  points~iii and~iv are not covered by MISRA~C:2012).
\item[7.1.a (R$\star$)] No variable shall have a name that is a keyword
  of C, C++, or any other well-known extension of the C programming language,
  including specifically K\&R C and C99.
  (MISRA~C:2012 Rule~21.2 covers only the keywords of the applicable
  C~standard).
\item[7.1.b (R$\star$)] No variable shall have a name that overlaps with
  a variable name from the C Standard Library
  (MISRA~C:2012 Rule~21.2 is more general).
\item[7.1.c (R)] No variable shall have a name that begins with an underscore
  (MISRA~C:2012 Rule~21.2 prevents declaration of any identifier with
  a reserved name, and this includes those beginning with underscore).
\item[7.1.d (R)] No variable name shall be longer than 31 characters
  (MISRA~C:2012 Rule~5.1 requires external identifiers to be distinct,
  which, for a generic C99 compiler, implies different external variable names
  shall differ in the first 31 characters).
\item[8.3.b (R)] All \verb+switch+ statements shall contain a \verb+default+
  block
  (MISRA~C:2012 Rule~16.5 prescribes presence and positioning of the the
  \verb+default+ label).
\item[8.3.c (R)] Any \verb+case+ designed to fall through to the next
  shall be commented to clearly explain the absence of the
  corresponding \verb+break+
  (MISRA~C:2012 Rule~16.3 requires presence of the final \verb+break+
  apart from the case of consecutive labels).
\item[8.4.b (R)] With the exception of the initialization of a loop
  counter in the first clause of a \verb+for+ statement and the change
  to the same variable in the third, no assignment shall be made in
  any loop's controlling expression
  (MISRA~C:2012 Rule~14.2 covers this rule, but is more general).
\item[8.4.c (R)] Infinite loops shall be implemented via controlling
  expression \verb+for (;;)+
  (MISRA~C:2012 Rule~14.2 allows \verb+for (;;)+ specifically for
  the purpose of expressing infinite loops).
\item[8.5.a (R)] The use of goto statements shall be restricted
  as per rule \textbf{1.7.c}
  (MISRA~C:2012 Rule~15.1, advisory ban of \verb+goto+,
  and Rule~15.2, required limitation to forward jumps).
\item[8.5.b (R)] C Standard Library functions \verb+abort()+, \verb+exit()+,
  \verb+setjmp()+, and \verb+longjmp()+ shall not be used
  (MISRA~C:2012 Rule~21.8 bans \verb+abort()+, \verb+exit()+ and other
  functions of \verb+<stdlib.h>+, MISRA~C:2012 Rule~21.4 bans all
  uses of \verb+<setjmp.h>+,
  including \verb+setjmp()+ and \verb+longjmp()+).
\end{STDdescription}

\subsection{BARR-C:2018 Guidelines with Related Guidelines in MISRA~C:2012}

10 guidelines of BARR-C:2018 have related, though not matching, counterpart
in MISRA~C:2012.
They are:

\begin{STDdescription}
\item[1.1.c (D)] Minimize the use of proprietary compiler language keyword
  extensions, \verb+#pragma+, and inline assembly
  (MISRA~C:2012 Rule~1.2 advises not to use language extensions;
  Dir~1.1 requires to document all relevant implementation-defined
  behaviors including those provided via new keywords and \verb+#pragma+
  directives; Dir~4.3 requires to encapsulate and isolate inline assembly
  code; Dir~4.2 advises to document all usage of assembly language).
\item[1.4.b (R)]
  Unless it is a single identifier or constant, each operand of the
  \verb+&&+ and \verb+||+ operators shall be surrounded by parentheses
  (compliance with this rule contributes to compliance
  with MISRA~C:2012 Rule~12.1).
\item[4.2.c (R)] The header file shall identify only the procedures,
  constants, and data types (via prototypes or macros, \verb+#define+,
  and \verb+typedef+s, respectively) about which it is strictly necessary
  for other modules to be informed
  (MISRA~C:2012 Dir~4.8 advises the use of opaque pointers for pointers
  to structures or unions that are never dereferenced within
  a translation unit).
\item[5.2.c (D$\star$)] Use of the keyword \verb+char+ shall be restricted
  to the declaration of and operations concerning strings
  (MISRA~C:2012 Rules~10.1, 10.2, and~10.4 restrict the use of plain
  \verb+char+ objects to the representation of characters and strings).
\item[5.4.b (D$\star$)] When floating point calculations are necessary:
  (i) Use the C99 type names \verb+float32_t+, \verb+float64_t+,
      and \verb+float128_t+.
  (ii) Append an `\verb+f+' to all single-precision constants.
  (iii) Ensure that the compiler supports double precision,
        if your math depends on it.
  (iv) Never test for equality or inequality of floating point values.
  (v) Always invoke the \verb+isfinite()+ macro to check that prior
      calculations have resulted in neither \verb+INFINITY+ nor \verb+NAN+.
  (MISRA~C:2012 Dir~4.6 covers point~i, Dir~1.1 covers point~iii).
\item[5.5.a (D$\star$)] Appropriate care shall be taken to prevent the compiler
  from inserting padding bytes within \verb+struct+ or \verb+union+ types
  used to communicate to or from a peripheral or over a bus or network
  to another processor.
  (MISRA~C:2012 Rule~19.2 advises not to use \verb+union+,
  Dir~1.1 requires to document all relevant implementation-defined behaviors,
  including those involving padding bytes).
\item[5.6.a (R)] Boolean variables shall be declared as type \verb+bool+
  (MISRA~C:2012 Rule~10.1 prevents uses of non-Booleans in Boolean-expecting
  contexts; MISRA~C:2012 also recommends using \verb+<stdbool.h>+
  when available \cite[Appendix~D.6.4]{MISRA-C-2012-Revision-1}).
\item[5.6.b (R)] Non-Boolean values shall be converted to Boolean via use
  of relational operators (e.g., \verb+<+ or \verb+!=+), not via casts
  (MISRA~C:2012 Rule~10.5 bans casting to Booleans, among other type casts).
\item[8.2.c (R$\star$)] Assignments shall not be made within an \verb+if+
  or \verb+else if+ test
  (MISRA~C:2012 Rule~13.4 advises not to use the result of assignment operators
  in any way; MISRA~C:2012 Rule~14.4 requires the controlling expression of
  \verb+if+ statements to have Boolean type).
\item[8.4.d (R)] Each loop with an empty body shall feature a set of braces
  enclosing a comment to explain why nothing needs to be done until after
  the loop terminates
  (MISRA~C:2012 Rule~15.6 requires the braces but not the comment).
\end{STDdescription}

\subsection{BARR-C:2018 Guidelines without Correspondence in MISRA~C:2012}
\label{sec:barr-c-2018-guidelines-without-correspondence}

17 guidelines of BARR-C:2018 are not stylistic and
have no counterpart in MISRA~C:2012.
The fact that this category is not empty, strictly speaking,
implies that the objective of making the BARR-C subset of C
a superset of MISRA~C:2012 has not been achieved by BARR-C:2018.
In order to fully understand the matter, it helps to divide
this category into further subcategories:

\subsubsection{Ban on Obsolete Keywords}

\begin{STDdescription}
\item[1.7.a (R)] The \verb+auto+ keyword shall not be used.
\item[1.7.b (R)] The \verb+register+ keyword shall not be used.
\end{STDdescription}

The \verb+auto+ keyword is only in the language for historical
reasons, as it serves no useful purpose.
To the contrary, it may be used to declare implicit \verb+int+
variables with declarations like \verb+auto x+.
While this violates a constraint of C99
\cite[Section~6.7]{ISO-C-1999-consolidated-TC3} (and thus Rule~1.1.a
of BARR-C:2018 and Rule~1.1 of MISRA~C:2012), many compilers
still generate code for that, with or without producing a warning.
Implicit \verb+int+ for C90 is banned by Rule~8.1 of MISRA~C:2012.
Summarizing, the only good reason to keep \verb+auto+ in a C~subset
is to accommodate legacy code.

The \verb+register+ keyword is also in the language for historical reasons:
since at least a couple of decades, compilers are much better than humans
in deciding which variables should be allocated to registers taking into
account the registers supply of the target processor.
It shares with \verb+auto+ the disadvantage of allowing
implicit \verb+int+ declarations like in \verb+register x+.
The only potentially interesting use case of \verb+register+ is
in preventing the taking of addresses of automatic variables: the declaration
\verb+register float y+ does not allow the address of \verb+y+
to be taken.  This could help in preventing undefined behavior caused
by dangling references: while MISRA~C:2012 has Rule 18.6 to prevent them,%
\footnote{``The address of an object with automatic storage shall not
  be copied to another object that persists after the first object has
  ceased to exist'' \cite[Rule~18.6]{MISRA-C-2012-Revision-1}.}
BARR-C has no guidelines to mitigate this risk.

\subsubsection{Development Process}

\begin{STDdescription}
\item[4.4.a (D)] A set of templates for header files and source files shall
  be maintained at the project level.
\item[5.5.b (D$\star$)] Appropriate care shall be taken to prevent the compiler
  from altering the intended order of the bits within bit-fields.
\item[6.5.a (D)] The compiler must be informed that the function is an ISR
  by way of a \verb+#pragma+ or compiler-specific keyword, such as
  ``\verb+__interrupt+''.
\item[6.5.c (D$\star$)] Ensure that ISRs are not inadvertently called from other
  parts of the software.
\item[6.5.d (D)] A stub or default ISR shall be installed in the vector
  table at the location of all unexpected or otherwise unhandled interrupt
  sources; each such stub could attempt to disable future interrupts
  of the same type.
\end{STDdescription}

These are prescriptions on the development process: when the compiler
does support them, they require the use  language extensions.
As such, they are not strictly related to language subsetting.

\subsubsection{Definite Language Subsetting Guidelines}

\begin{STDdescription}
\item[1.7.d (D)] It is a preferred practice to avoid all use of the
  \verb+continue+ keyword.
\item[1.8.c (D$\star$!)] The \verb+volatile+ keyword shall be used
  whenever appropriate.
\item[4.2.a (R)] There shall always be precisely one header file for each
  source file and they shall always have the same root name.
\item[4.2.d (D)] No public header file shall contain a \verb+#include+
  of any private header file.
\item[5.2.b (R$\star$)] The keywords \verb+short+ and \verb+long+ shall
  not be used.
\item[5.4.a (D)] Avoid the use of floating point constants and variables
  whenever possible.
\item[7.2.d (R)] Any pointer variable lacking an initial address shall be
  initialized to \verb+NULL+.
\item[8.4.a (D)] Magic numbers shall not be used as the initial value or
  in the endpoint test of a \verb+while+, \verb+do+\dots\verb+while+,
  or \verb+for+ loop.
\item[8.1.a (R$\star$!)] A declaration shall not declare more than one
  declarator.\footnote{BARR-C:2018 \cite{BARR-C-2018} wording
    erroneously mentions the comma operator (which is a different
    thing) not to be used within variable declarations. The intention,
    however, is clearly the one to avoid \verb+char * x, y+
    as well as \verb+int *f(void), g(int z)+.}
\item[8.2.b (R)] Nested \verb+if+ \dots \verb+else+ statements shall
  not be deeper than two levels.
\end{STDdescription}

Concerning 1.7.d, the \verb+continue+ statement was banned
by Rule~14.5 of MISRA~C:2004 \cite{MISRA-C-2004}; it is allowed
without restrictions in MISRA~C:2012.
For 1.8.c, casts removing \verb+const+ or \verb+volatile+ qualification
are banned by MISRA~C:2012 Rule~11.8; omitting the \verb+volatile+
qualification is a source of bugs that may be very difficult to diagnose.
Regarding the advice of 5.4.a not to use floating point constants and
variables unless necessary, MISRA~C:2012 Dir~1.1 requires documentation
and understanding of, among other things, the many implementation-defined
aspects of floating point arithmetic, when used.
Finally, the lack of a guideline in MISRA~C:2012 that, similarly
to BARR-C:2018's 8.1.a, recommends against having more than one declarator
per declaration, is probably due to an oversight.

\section{BARR-C:2018 Stylistic Guidelines}
\label{sec:barr-c-2018-stylistic-guidelines}

The matter of style, while being essential to ensure program readability,
is highly subjective.  Everyone in software development knows that
matters apparently as futile as the ``right'' indent size has the
potential of causing friction within the development team.
Nonetheless, as observed in \cite{BARR-C-2018}, ``[individual] programmers
do not own the software they write.  All software development is work
for hire for an employer or a client [...]''.  So, someone has to make
stylistic choices and consistency is usually much more important than
the details of the chosen rules.
Readers interested in programming style are referred to the classic
{\it The Elements of Programming Style} \cite{KernighanP78},
published in 1978 with examples in PL/I and Fortran, but still a source
of good advice, which is largely independent from the programming language.

79 out of 143 guidelines of BARR-C:2018 \cite{BARR-C-2018} are stylistic in
nature.  They cover guidance on line width, horizontal spacing
(blanks spaces, tabs, alignment, indentation), vertical spacing (new-line
and other control characters), further code layout issues, naming (modules
and files, types, functions variables), language, comments,
and source file contents.  They are illustrated in the following sections.

\subsection{Line Width Guidance}

BARR-C:2018 has one rule concerning the maximum line width
for program sources.

\begin{STDdescription}
\item[1.2.a (R)] Limit the length of all lines in a program
  to a maximum of 80 characters.
\end{STDdescription}

The rationale for this rule is to increase readability.
While the limitation to 80 characters originates from the width
of IBM punch cards and 80-column-wide screens, very
long lines are difficult to read on a computer screen
\cite{BergfeldMillsW87}.
Longer lines might be broken by the editor in ways that impair
reading further or, worse, the final part of the line might
be shown in a way that escapes the reader's attention.

Code reviews can benefit from the availability
of high-quality printed listings, which are usually limited
to 65--70 characters per line, depending on the font and paper size.
For maximum readability, the majority of program text should fit
into 55 characters \cite{Covington94}.
Depending on the technology used to obtain the printed listing, the
final part of long lines may simply be not printed.

\subsection{Horizontal Spacing Guidance}

It is well known that the systematic use of horizontal space
helps readability.  A balance has to be found between opposite
goals: keep things separated enough to avoid clutter, keep them
close enough to convey the connection between them; indent enough
to make inclusions noticeable, but not too much to avoid long lines
or line splits; aligning can improve readability, but overdoing it
might impair readability.

\subsubsection{Blank Spaces Guidance}

BARR-C:2018 has 13 rules concerning the presence or absence of blank spaces,
namely:

\begin{STDdescription}
\item[3.1.a (R)] Use one space after the keywords \verb+if+, \verb+while+,
  \verb+for+, \verb+switch+, and \verb+return+.
\item[3.1.b (R)] Use one space before and after assignment operators
  \verb#=#, \verb#+=#, \verb#-=#, \verb#*=#, \verb#/=#, \verb#%=#,
  \verb#&=#, \verb#|=#, \verb#^=#, \verb#~=#, and \verb#!=#.
\item[3.1.c (R)] Use one space before and after binary operators
  \verb#+#, \verb#-#, \verb#*#, \verb#/#, \verb#%#, \verb#<#, \verb#<=#,
  \verb#>#, \verb#>=#, ==,\verb#!=#, \verb#<<#, \verb#>>#, \verb#&#,
  \verb#|#, \verb#^#, \verb#&&#, and \verb#||#.
\item[3.1.d (R)] Use no space on the operand side for unary operators
  \verb#+#, \verb#-#, \verb#++#, \verb#--#, \verb#! #, and \verb#~#.
\item[3.1.e (R)] Use one space before and after pointer operators \verb|*|
  and \verb|&| in declarations, no space on the operand side in other contexts.
\item[3.1.f (R)] Use one space before and after \verb|?| and \verb|:|
  characters comprising the ternary operator.
\item[3.1.g (R)] Use no spaces around the structure pointer (\verb+->+)
  and structure member (\verb+.+) operators.
\item[3.1.h (R)] Use no spaces around square brackets of the array subscript
  operator \verb+[]+ unless required by another rule.
\item[3.1.i (R)] Use no spaces after \verb+(+ or before \verb+)+ in
  expressions.
\item[3.1.j (R)] Use no spaces around \verb+(+ and \verb+)+ in function calls,
  one space between the function name and \verb+(+ in function definitions.%
\footnote{BARR-C-2018 says \emph{declarations} but such space is
  not present in the non-defining function declarations presented
  in \texttt{crc.h} \cite[Appendix~D]{BARR-C-2018}.}
\item[3.1.k (R)] Use one space after each comma separating function parameters.
\item[3.1.l (R)] Use one space after each semicolon separating the \verb+for+
  statement clauses.
\item[3.1.m (R)] Use no space before the semicolon terminating a statement.
\end{STDdescription}

\subsubsection{Tab Guidance}

\begin{STDdescription}
\item[3.5.a (R)] Do not use the tab character (ASCII \verb+HT+, \verb+0x09+).
\end{STDdescription}

Tab cannot be expected to be consistently set across editors and browsers.
In addition, mixing tabs and spaces is problematic as far as searches and
substitutions are concerned.

\subsubsection{Alignment Guidance}

BARR-C:2018 has 5 rules concerning alignments that, emphasizing similarity,
improve readability:

\begin{STDdescription}
\item[3.2.a (R)] Align names of variables within a series of declarations.
\item[3.2.b (R)] Align names of \verb+struct+ and \verb+union+ members.
\item[3.2.c (R)] Align assignment operators within a block of adjacent
  assignment statements.
\item[3.2.d (R)] Align the leading \verb|#| in preprocessor directives
\item[8.3.a (R$\star$)] The \verb+break+ for each \verb+case+ shall be
  indented to align with the associated \verb+case+.
\end{STDdescription}

\subsubsection{Indentation Guidance}

BARR-C:2018 has 2 rules and 1~directive concerning indentation,
namely:

\begin{STDdescription}
\item[3.4.a (R)] Use 4 character per indentation level.
\item[3.4.b (R)] Case labels shall be aligned and the contents of each
  case block shall be indented once from there.
\item[3.4.c (D)] When long lines are broken into multiple lines,
  use indentation to maximize readability.
\end{STDdescription}

The directive recommends to choose the breaking point wisely, e.g.,
to facilitate the interpretation of logical source lines of code
containing operators, and to use indentation to emphasize the
continuation context.

\subsection{Vertical Spacing Guidance}

\subsubsection{New-lines Guidance}

BARR-C:2018 has 3 rules concerning the use of new-line characters
to increase readability, namely:

\begin{STDdescription}
\item[3.3.a (R)] Write at most one statement per line.
\item[3.3.b (R)] A blank line shall precede each natural block of code
  (loops, conditionals, switches, consecutive declarations).
\item[3.3.c (R)] Each source file shall be terminated by a comment marking
  the end of file followed by a blank line.
\end{STDdescription}

It is worthwhile noting that rule \textbf{3.3.c} also avoids a case of
undefined behavior: this happens when a source file that is not empty
does not end in a new-line character
\cite[5.1.1.2p2]{ISO-C-1999-consolidated-TC3}.

\subsubsection{Control Characters Guidance}

Given that the horizontal tab \verb+HT+ is forbidden by
rule \textbf{3.5.a}, very few ASCII control character are
allowed by \cite{BARR-C-2018}:

\begin{STDdescription}
\item[3.6.a (R)] End source code lines with \verb+LF+ (ASCII \verb+0x0A+),
  not with the pair \verb+CR+-\verb+LF+ (ASCII \verb+0x0D+ \verb+0x0A+).
\item[3.6.b (R)] Do not use other control characters apart from
  the form feed character \verb+FF+ (ASCII \verb+0x0C+).
\item[6.2.b (D)] Whenever possible, all functions shall be made to start
  at the top of a printed page, except when several small functions can fit
  onto a single page.
\end{STDdescription}

\subsection{Mixed Code Layout Guidance}

\begin{STDdescription}
\item[1.3.b (R)] Blocks shall be delimited by a left brace (\verb+{+)
  alone in the line and a right brace (\verb+}+) alone in the line
  and in the same column as the left brace.
\item[6.2.a (D)] All reasonable effort shall be taken to keep the length
  of each function limited to one printed page, or a maximum of 100 lines.
\item[7.2.b (R)] Define local variables as you need them, rather than all
  at the top of a function.
\item[7.2.c (D)] If project- or file-global variables are used, their
  definitions shall be grouped together and placed at the top of a
  source code file.
\item[8.2.a (R)] It is a preferred practice that the shortest
  (measured in lines of code) of the \verb+if+ and \verb+else if+
  clauses should be placed first.
\item[8.6.a (R$\star$)] When evaluating the equality of a variable against
  a constant, the constant shall always be placed to the left of the
  equal-to operator (\verb+==+).
\end{STDdescription}

\subsection{Naming Guidance}

One of the crucial activities in software development is choosing the
right names.  The interested reader can find detailed guidance on naming
in \cite{LedgardT87} and \cite[p.\ 104 ff.]{KernighanP99}.

\subsubsection{Module and File Names}

In \cite{BARR-C-2018} a \emph{module} is a logical entity with a name.
A module is implemented in one header file and one source file.

\begin{STDdescription}
\item[4.1.a (R$\star$)] Module names shall consist entirely of lowercase
  letters, numbers, and underscores.
\item[4.1.b (R)] Module names shall be unique in their first 8 characters
  and end with suffices \verb+.h+ and \verb+.c+ for the header and source
  file names, respectively.
\item[4.1.c (R$\star$)] No module's header file name shall share the name
  of a header file from the C or C++ standard libraries.
\item[4.1.d (R)] Modules containing a \verb+main()+
  function shall have the word
  ``\verb+main+'' as part of their source file name.
\end{STDdescription}

\subsubsection{Type Names}

\begin{STDdescription}
\item[5.1.a (R)] The names of all new data types, including structures,
  unions, and enumerations, shall consist only of lowercase characters
  and internal underscores and end with \verb+_t+.
\item[5.1.b (R)] All new structures, unions, and enumerations shall be named
  via a \verb+typedef+.
\item[5.1.c (D)] The name of all public data types shall be prefixed
  with their module name and an underscore.
\end{STDdescription}

\subsubsection{Function Names}

\begin{STDdescription}
\item[6.1.e (R)] No function name shall contain any uppercase letters.
\item[6.1.f (R)] No macro name shall contain any lowercase letters.
\item[6.1.g (R)] Underscores shall be used to separate words in procedure names.
\item[6.1.h (D)] Each procedure's name shall be descriptive of its purpose.
\item[6.1.i (R)] The names of all public functions shall be prefixed with their module name and an underscore (e.g., \verb+sensor_read()+).
\item[6.4.a (D)] All functions that encapsulate threads of execution
  (a.k.a., tasks, processes) shall be given names ending with
  ``\verb+_thread+'' (or ``\verb+_task+'', ``\verb+_process+'').
\item[6.5.b (D)] All functions that implement ISRs shall be given names
  ending with ``\verb+_isr+''.
\end{STDdescription}

\subsubsection{Variable Names}

\begin{STDdescription}
\item[7.1.e (R)] No variable name shall be shorter than 3 characters.
\item[7.1.f (R$\star$)] No variable name shall contain any uppercase letters.
\item[7.1.g (R)] No variable name shall contain any numeric value that
  is called out elsewhere.
\item[7.1.h (R)] Underscores shall be used to separate words in variable names.
\item[7.1.i (D)] Each variable's name shall be descriptive of its purpose.
\item[7.1.j (R$\star$)] The names of any global variables shall begin with the
  letter `\verb+g+'.
\item[7.1.k (R$\star$)] The names of any pointer variables shall begin with the
  letter `\verb+p+'.
\item[7.1.l (R)] The names of any pointer-to-pointer variables shall begin
  with the letters `\verb+pp+'.
\item[7.1.m (D$\star$)] The names of all integer variables containing Boolean
  information (including 0 vs. non-zero) shall begin with the letter
  `\verb+b+' and phrased as the question they answer.
\item[7.1.n (R)] The names of any variables representing non-pointer handles
  for objects, e.g., file handles, shall begin with the letter `\verb+h+'.
\item[7.1.o (R)] In the case of a variable name requiring multiple
  of the above prefixes, the order of their inclusion before
  the first underscore shall be \verb+[g][p|pp][b|h]+.
\end{STDdescription}

\subsection{Language Guidance}

\begin{STDdescription}
\item[1.5.a (D)] Use only common and widely understood abbreviations
  and acronyms.
\item[1.5.b (D)] Maintain a project-specific table of abbreviations
  and acronyms.
\end{STDdescription}

\subsection{Comment Guidance}

\begin{STDdescription}
\item[1.6.a (D)] comment each cast describing how the code ensures
  proper behavior.
\item[2.1.a (D)] Single-line comments in the C++ style
  (i.e., preceded by \verb+//+) are a useful and acceptable alternative
  to traditional C style comments (i.e., \verb+/*+ \dots \verb+*/+).
\item[2.2.a (D)] All comments shall be written in clear and complete
  sentences, with proper spelling and grammar and appropriate punctuation.
\item[2.2.b (D)] The most useful comments generally precede a block of code
  that performs one step of a larger algorithm.
\item[2.2.c (D)] Avoid explaining the obvious.
\item[2.2.d (D)] The number and length of individual comment blocks shall
  be proportional to the complexity of the code they describe.
\item[2.2.e (D)] Whenever a comment references an external document,
  the comment shall include sufficient references to the original source.
\item[2.2.f (D)] Whenever a flow chart or other diagram is needed to
  sufficiently document the code, the drawing shall be maintained with
  the source code under version control and the comments should
  reference the diagram by file name or title.
\item[2.2.g (D$\star$)] All assumptions shall be spelled out in comments.
\item[2.2.h (R)] Each module and function shall be commented in a manner
  suitable for automatic documentation generation, e.g., via Doxygen.
\item[2.2.i (D$\star$)] Use the following capitalized comment markers
  to highlight important issues:
  (i) ``\verb+WARNING:+'' alerts a maintainer there is risk in changing
  this code.
  (ii) ``\verb+NOTE:+'' provides descriptive comments about the ``why''
  of a chunk
  of code---as distinguished from the ``how'' usually placed in comments.
  (iii) ``\verb+TODO:+'' indicates an area of the code is still under
  construction and explains what remains to be done.
\end{STDdescription}

\subsection{Source File Contents Guidance}

\begin{STDdescription}
\item[4.3.a (D)] Each source file shall include only the behaviors
  appropriate to control one ``entity''.
\item[4.3.b (R)] Each source file shall be comprised of some or all of the
  following sections, in the order listed: comment block; include statements;
  data type, constant, and macro definitions; static data declarations;
  private function prototypes; public function bodies;
  then private function bodies.
\item[4.3.c (R$\star$)] Each source file shall always \verb+#include+
  the header file of the same name (e.g., file \verb+adc.c+
  should \verb+#include "adc.h"+), to allow the compiler to confirm
  that each public function and its prototype match.
\item[4.3.d (R)] Absolute paths shall not be used in include file names.
\item[4.3.e (R)] Each source file shall be free of unused include files.
\item[4.3.f (R)] No source file shall \verb+#include+ another source file.
\end{STDdescription}

\section{Adoption of BARR-C:2018 and MISRA~C:2012}
\label{sec:adoption-of-barr-c-2018-and-misra-c-2012}

Given the substantial compatibility of BARR-C:2018 and MISRA~C:2012,
they can both
be applied, at least partially.  They have in common least a few
important characteristics:
\begin{enumerate}
\item
  they are established and rather well known in the embedded systems'
  community (even though MISRA~C predates BARR-C of more than a decade):
\item
  they have both been designed with static analysis in mind (even though
  the MISRA~C:2012 guidelines are more precisely specified than the
  BARR-C:2018 ones);
\item
  they both support a deviation process.
\end{enumerate}

For safety-related projects, the adoption of the stylistic subset of
BARR-C:2018 can be part of complying with the spirit of MISRA~C:2012.
In fact
\cite[Section 5.2.2, \emph{Process activities expected by MISRA C}]{MISRA-C-2012-Revision-1}:
\begin{quote}
It is recognized that a consistent style assists programmers in
understanding code written by others. However, since style is a matter
for individual organizations, MISRA C does not make any
recommendations related purely to programming style. It~is expected
that local style guides will be developed and used as part of the
software development process.
\end{quote}
While the stylistic guidance provided by BARR-C:2018 may not suit the taste
of everyone, the fact that it exists, is publicly available, and is
supported by tools is a strong point in its favor.
Another aspect to be taken into account is that BARR-C:2018 is flexible
as far as quantities are concerned
\cite[\emph{Deviation Procedure}]{BARR-C-2018}:
\begin{quote}
At the project level, rules that indicate a specific quantity of
something (e.g., the number of characters per indent or maximum lines
in a function) can be changed to enforce a different quantity that
works better in the actual development tools. The specific quantity
is not typically the key property of these types of rules.
\end{quote}

There is another important way in which BARR-C and MISRA~C can coexist:
by providing a smooth entry path for organizations and projects that
do not have (yet) or only have partial requirements about MISRA~C
compliance.
In fact, BARR-C:2018 fills an important gap: there is way too much
C~software that, in the absence of normative or contractual obligations
to comply with mature coding standards such as the MISRA ones, is
developed in unconstrained~C and not subjected to any static analysis.
For such projects, moving from the wild to (partial) compliance with
BARR-C:2018 would constitute an important step forward, and one that
is not toot difficult to make.
The further step would be, at least for the critical project, to
move from BARR-C:2018 compliance to MISRA~C:2012 compliance.
This step would be significantly easier to take, compared to the
case where the starting point is ``no coding standard, no static analysis.''
For a least two reasons:
\begin{enumerate}
\item
  Culture: a team already trained to the use of coding standards and
  static analysis can more easily move to a more complex coding standard
  and static analyses.
\item
  Starting point: part of the work required to comply with
  MISRA~C:2012 has already been done by complying to BARR-C:2018.
\end{enumerate}

A project that is compliant with BARR-C:2018 will be mostly compliant
with 22 MISRA~C:2012 guidelines.  These are:\footnote{Mandatory MISRA~C:2012
  guidelines are set in boldface, advisory ones in italics.}
\textit{Dir~4.4} (by \textbf{2.1.c}),
\textit{Dir~4.6} (by \textbf{5.2.a} and \textbf{5.4.b}),
\textit{Dir~4.9} (by \textbf{6.3.a}),
Dir~4.10 (by \textbf{4.2.b}),
Rule~1.1 (by \textbf{1.1.a} and \textbf{1.1.b}),
Rule~3.1 (by \textbf{2.1.b}),
Rule~3.2 (by \textbf{2.1.b}),
Rule~6.1 (by \textbf{5.3.a}),
Rule~7.2 (by \textbf{5.3.c}),
Rule~8.2 (by \textbf{6.2.f}),
\textit{Rule~8.7} (by \textbf{1.8.a}),
Rule~8.8 (by \textbf{6.2.e}),
\textit{Rule~8.13} (by \textbf{1.8.b}),
\textbf{Rule~9.1} (by \textbf{7.2.a}.
\textit{Rule~12.1} (by \textbf{1.4.a}),
\textit{Rule~15.1} (by \textbf{1.7.c} or \textbf{8.5.a}),%
\footnote{\label{goto-footnote}The BARR-C:2018 goto rules
  cover both MISRA~C:2012 Rule~15.1 and Rule~15.2 if no gotos
  are allowed but only Rule~15.2 if forward gotos are allowed.}
Rule~15.2 (by \textbf{1.7.c} or \textbf{8.5.a}),%
\footnote{See footnote~\ref{goto-footnote}.}
\textit{Rule~15.5} (by \textbf{6.2.c}),
Rule~15.6 (by \textbf{1.3.a}),
Rule~15.7 (by \textbf{8.2.d}),
Rule~20.4 (by \textbf{1.1.d}),
Rule~20.7 (by \textbf{6.3.b}.

In addition, the same project will be partially compliant to the
following 19 guidelines:
Dir~1.1 (by \textbf{1.1.c}),
\textit{Dir~4.2} (by \textbf{1.1.c}),
Dir~4.3 (by \textbf{1.1.c}),
\textit{Dir~4.8} (by \textbf{4.2.c}),
\textit{Rule~1.2} (by \textbf{1.1.c}),
Rule~5.1 (by \textbf{6.1.d} and \textbf{7.1.d}),
Rule~8.5 (by \textbf{6.2.d}),
Rule~10.1 (by \textbf{5.2.c} and \textbf{5.3.b} \textbf{5.6.a}),
Rule~10.2 (by \textbf{5.2.c}),
Rule~10.4 (by \textbf{5.2.c} and \textbf{5.3.c}),
\textit{Rule~10.5} (by \textbf{5.6.b}),
\textit{Rule~13.4} (by \textbf{8.2.c}),
Rule~14.2 (by \textbf{8.4.b} and \textbf{8.4.c}),
Rule~16.1 (by \textbf{1.3.a}),
Rule~16.5 (by \textbf{8.3.b}),
\textit{Rule~19.2} (by \textbf{5.5.a}),
Rule~21.2 (by \textbf{6.1.a}), \textbf{6.1.b}), \textbf{6.1.c} and \textbf{7.1.a}),
Rule~21.4 (by \textbf{8.5.b}),
Rule~21.8 (by \textbf{8.5.b}).

Passing from BARR-C:2018 compliance to MISRA~C:2012 compliance
requires taking into account 9 further directives (7 required and 2 advisory)
and 123 further MISRA~C:2012 rules (13 mandatory, 87 required and 23 advisory).
Such guidelines are listed in the following, divided into broad categories.

\subsection{Prevention of undefined and unspecified behavior}
This category contains 5 directives and 61 rules:
  Dir~2.1,
  Dir~4.1,
  Dir~4.11,
  Dir~4.12,
  Dir~4.14,
  Rule~1.3,
  Rule~5.2,
  Rule~5.4,
  Rule~7.4,
  Rule~8.6,
  Rule~8.3,
  Rule~8.4,
  Rule~8.10,
  Rule~8.14,
  Rule~9.2,
  Rule~9.4,
  Rule~10.3,
  Rule~11.1,
  Rule~11.2,
  Rule~11.3,
  \textit{Rule~11.4},
  \textit{Rule~11.5},
  Rule~11.6,
  Rule~11.7,
  Rule~11.8,
  Rule~12.2,
  Rule~13.1,
  Rule~13.2,
  \textit{Rule~13.3},
  \textbf{Rule~13.6},
  Rule~17.1,
  \textbf{Rule~17.3},
  \textbf{Rule~17.4},
  \textbf{Rule~17.6},
  Rule~18.1,
  Rule~18.2,
  Rule~18.3,
  Rule~18.6,
  Rule~18.7,
  Rule~18.8,
  \textbf{Rule~19.1},
  \textit{Rule~20.1},
  Rule~20.2,
  Rule~20.3,
  Rule~20.6,
  \textit{Rule~20.10},
  Rule~20.11,
  Rule~21.1,
  Rule~21.3,
  Rule~21.5,
  Rule~21.6,
  Rule~21.7,
  Rule~21.9,
  Rule~21.10,
  Rule~21.11,
  \textit{Rule~21.12},
  \textbf{Rule~21.13},
  Rule~21.14,
  Rule~21.16,
  \textbf{Rule~21.17},
  Rule~21.18,
  \textbf{Rule~21.19},
  \textbf{Rule~22.2},
  \textbf{Rule~22.4},
  \textbf{Rule~22.5},
  \textbf{Rule~22.6}.

\subsection{Limiting dependence on implementation-defined behavior}
The following 2 rules help with program portability by avoiding
some implementation-defined behaviors:
  Rule~4.1,
  Rule~22.3.

\subsection{Enhancing readability}
There are 1 directive and 21 rules in this category:
  \textit{Dir~4.5}
  \textit{Rule~2.3},
  \textit{Rule~2.4},
  \textit{Rule~2.5},
  \textit{Rule~2.6},
  \textit{Rule~2.7},
  \textit{Rule~4.2},
  Rule~5.3,
  Rule~7.3,
  \textit{Rule~8.9},
  Rule~8.12,
  Rule~9.3,
  Rule~9.5,
  Rule~11.9,
  \textit{Rule~12.3},
  \textit{Rule~15.4},
  Rule~16.2,
  Rule~16.6,
  Rule~16.7,
  \textit{Rule~18.5},
  \textit{Rule~20.5},
  Rule~20.14.

\subsection{Enhancing verifiability}
These guidelines help the analysis and verifiability of the source code.
There are 1 such directive and 5 rules:
  \textit{Dir~4.13},
  \textit{Rule~8.11},
  Rule~15.3,
  Rule~16.4,
  \textit{Rule~17.5},
  Rule~17.7.

\subsection{Reducing developer confusion}
These 25 rules avoid code that may lead to developer confusion:
  Rule~2.1,
  Rule~2.2,
  Rule~5.5,
  Rule~5.6,
  Rule~5.7,
  Rule~5.8,
  \textit{Rule~5.9},
  Rule~6.2,
  Rule~7.1,
  Rule~8.1,
  Rule~10.6,
  Rule~10.7,
  Rule~10.8,
  \textit{Rule~12.4},
  \textbf{Rule~12.5},
  Rule~13.5,
  Rule~14.4,
  Rule~16.3,
  \textit{Rule~17.8},
  \textit{Rule~18.4},
  Rule~20.8,
  Rule~20.9,
  Rule~20.12,
  Rule~20.13,
  Rule~21.15.

\subsection{Prevention of unexpected run-time behavior}
These guidelines help prevent run-time failures and unexpected results.
These are 2 directives and 9 rules in this category:
  Dir~3.1,
  Dir~4.7,
  Rule~14.1,
  Rule~14.3,
  Rule~17.2,
  \textbf{Rule~21.20},
  Rule~22.1,
  Rule~22.7,
  Rule~22.8,
  Rule~22.9,
  Rule~22.10.

\section{Conclusion}
\label{sec:conclusion}

In this paper we have illustrated the connections between the two
most widely used coding standards in the embedded systems industry:
MISRA~C and BARR-C.
We have briefly recalled some of the advantages and disadvantages of
using the C~programming language for embedded systems and how its
uncontrolled use is not adequate for the development of systems that
are even moderately critical.
We have also summarized the background, motivation and history
of the MISRA project and of MISRA~C in particular.

After recalling the main features of the MISRA~C:2012 guidelines,
we have introduced the BARR-C:2018 coding standard from the MISRA
point of view.  All BARR-C:2018 guidelines have been presented, divided
into two broad categories: those dealing with language subsetting and project
management, and those concerning programming style.
The BARR-C:2018 guidelines in the first category have been further classified
on the basis of their overlap with the MISRA~C:2012 guidelines.
Those in the second category have been further classified into subcategories
of stylistic guidance.

We have then explained the potential synergy between BARR-C:2018
and MISRA~C:2012.  They are amenable to \emph{parallel adoption}:
a project seeking MISRA~C compliance can use the coding style
portion of BARR-C:2018 thereby satisfying that MISRA~C recommendation
of adopting and enforcing a consistent coding style.
They are also amenable to \emph{serial adoption}: when a MISRA~C
compliance requirement is not (yet) present, the adoption of BARR-C:2018
is a major improvement with respect to the situation where no coding standards
and no static analysis are used.
We have shown that complying with BARR-C:2018 entails compliance
with a non-negligible subset of MISRA~C:2012.
While going from BARR-C:2018 compliance to MISRA~C:2012 compliance
still requires an effort that should not be underestimated,
BARR-C:2018 compliant projects and teams trained to its use and enforcement
are in very good position to tackle MISRA~C:2012 compliance of that
and other projects.

\subsection*{Acknowledgments}

For the notes on the history of MISRA and MISRA~C we are indebted
to Andrew Banks (LDRA, current Chairman of the MISRA C Working Group)
and David Ward (HORIBA MIRA, current Chairman of the MISRA Project).
We are also grateful to the following BUGSENG collaborators:
Abramo Bagnara, for the many discussions we had on the subject;
Anna Camerini for the composition of Figure~\ref{fig:misra-c-history}.


\providecommand{\noopsort}[1]{}

\end{document}